\documentstyle[12pt]{article}
\newcommand{\Sop}{{\cal S}}
\newcommand{\est}{{\bf X}}

\begin{document}
\baselineskip=24pt

\title{Broad Histogram Method for Multiparametric Hamiltonians}

\author{A. R. Lima\cite{email}, P. M. C. de Oliveira and T. J. P. Penna\\
Instituto de F\'{\i}sica, Universidade Federal Fluminense \\ 
Av. Litor\^anea, s/n - 24210-340 Niter\'oi, RJ, Brazil} 

\date{Last correction: December 7, 1999 - Printed: \today}
\maketitle

\begin{abstract}
  We extended the Broad Histogram Method in order to obtain spectral
 degeneracies for systems with multiparametric Hamiltonians. As
 examples we obtained the critical lines for the square lattice Ising
 model with nearest and next-nearest neighbor interactions and the
 antiferromagnetic Ising model in an external field. For each system,
 the entire critical line is obtained using data from a single
 computer run. We also discuss the accuracy and efficiency of our
 method.
\end{abstract}

\bigskip

The search for new and more efficient methods for computer simulations is
always an intensive task in science. Multispin coding techniques (see
\cite{pmcobook} and references therein), cluster algorithms
\cite{coniglio80,swendsen92,wolff89}, reweighting procedures
\cite{salzburg59,ferrenberg,marinari92} and methods which calculate
directly the spectral degeneracy \cite{berg91,lee93,hesselbo95,pmco96} are
a few remarkable examples (see \cite{marinari96,defelicio96,newman99} for
reviews). An efficient implementation of these methods can lead to an
enormous increase in speed and accuracy of computer simulations. 
Multiparametric simulation is another way to further improve the
efficiency. Multiparametric simulations allow, from a single computer run,
to explore the whole space of parameters defining a given system, instead
of repeating the whole process every time some parameter is changed.

For single-parameter simulations, several methods such as the
histogram method \cite{salzburg59}, simulated
tempering \cite{marinari92} , multicanonical ensemble
\cite{berg91,lee93}, $1/k$-sampling \cite{hesselbo95}, broad
histogram \cite{pmco96} among others have been shown to be successful.
On the other hand, only a few works in the literature treat
multiparametric simulations using such methods. The histogram method
\cite{ferrenberg} is easily generalized to any number of parameters
of the Hamiltonian. However the main problem in this approach is that
the number of simulations needed to cover a given region in the space
of parameters increases with the system size, in a $d$-dimensional
space, as $L^{d/2}$ for each parameter \cite{ferrenberg}. Shteto {\em
et al.} used the multicanonical method (Entropic Sampling
formulation) in order to study relaxation paths in the simulation of
an Ising model in an external field \cite{shteto97}. In that work
they pointed out that the entropic sampling can be generalized to any
number of parameters.  Although easily generalizable, a simple
implementation of this method has been shown to display a very slow
convergence \cite{newman99,lima99b}. On the other hand, methods such
as the traditional multicanonical formulation \cite{berg91} seem to
be not easily extendable to perform simulations with more than one
parameter. Recently a re-formulation of the broad histogram method
was introduced including an external field as a second parameter
\cite{kastner99}.

In this paper we perform a spectral degeneracy calculation by
applying the Broad Histogram method (BHM) to multiparametric
Hamiltonians. We show that the method can be generalized to any
number of parameters and the full range of them can be exploited. BHM
was introduced three years ago \cite{pmco96} and it is based on an
exact relation between the spectral degeneracy and some special
macroscopic quantities defined within the method itself. A remarkable
feature of BHM is its generality, since it is not restricted to any
specific dynamical rule: the macroscopic quantities needed can be
obtained by different procedures. BHM has been applied to a variety
of magnetic systems with accurate results in a very efficient way 
\cite{pmco96,lima99b,kastner99,lima99,pmco98,pmco98b,pmco99,wang98,munoz98}.

Any statistical system can be described by two different classes of
parameters: the ones which characterize it and are defined in the
Hamiltonian, and the others which control the interaction with the
environment. Let us consider the Hamiltonian of a given system as
composed by $P$ independent interaction terms %
\begin{equation}
  {\cal H}(\est) = \gamma_1 \Sop_{1}(\est) +  \gamma_2 \Sop_{2}(\est) + \dots + \gamma_{P} \Sop_{P}(\est)
\end{equation}
where the terms $\Sop_{1},..., \Sop_{P}$ depends only on the
microstate $\est$ of the system. $\gamma_1,...,\gamma_{P}$ are the
parameters associated to these terms and all together define the
complete Hamiltonian.  For instance, $\Sop_1$ could represent the
coupling with an external field, $\Sop_2$ a first-neighbor
interaction, and so on. The temperature, on the other hand, belongs
to the class of parameters which determines the interaction with the
environment.

Before presenting the method, it is fundamental to introduce the
concept of macrostate. To a given state $\est$ of the system, each
$\Sop_i(\est)$ gives a value $s_i$. These values
define a vector ${\bf s} = (s_1, s_2, \dots, s_{P})$.
Hence, the set of all states $\est$ corresponding to the same ${\bf
s}$ is called a {\em macrostate}.  The space defined by all
macrostates is called {\em space of macrostates}.  The interaction
with the environment corresponds to the choice of the statistical
ensemble, described by a weight function $f({\cal H},
\alpha_1,...,\alpha_{P_f}) \equiv f(\gamma_1,...,\gamma_{P},
\alpha_1,...,\alpha_{P_f}, {\bf s})$ whose functional form is known
{\em a priori}. The control parameters for such interaction are
$\alpha_1,...,\alpha_{P_f}$. Once the system (characterized by ${\bf
s},\gamma_1,...,\gamma_{P}$) and its interaction with the environment
(characterized by $\alpha_1,...,\alpha_{P_f}$) are defined, the
thermodynamic average of any quantity $Q$, can be written as %
\begin{equation}
  \langle Q \rangle_{\alpha_1,...,\alpha_{P_f}} = \frac{\sum_{\bf
s}{g({\bf s})\langle Q({\bf s}) \rangle f(\gamma_1,...,\gamma_{P},
\alpha_1,...,\alpha_{P_f},{\bf s})}} {\sum_{\bf s}{g({\bf
s})f(\gamma_1,...,\gamma_{P}, \alpha_1,...,\alpha_{P_f}, {\bf s})}}
\end{equation}
where $g({\bf s})$ is the spectral degeneracy, i.e. the number of
possible microstates of the system corresponding to the same
macrostate ${\bf s}$.  $\langle Q({\bf s}) \rangle$ is the uniform
average of $Q$ in the macrostate ${\bf s}$.  The parameters
introduced by $f$ are irrelevant for the definition of ${\bf s}$ and
consequently to the calculation of $g({\bf s})$ (this idea has
already been used in another context \cite{lima99,salazar99}).  Also,
$g({\bf s})$ is independent of the particular values
($\gamma_1,...,\gamma_{P}$) of the Hamiltonian parameters. Therefore,
if one is able to provide the spectral degeneracy $g({\bf s})$ and
$\langle Q({\bf s}) \rangle$, all information about the observable
$Q$ can be obtained for the full range of the parameters
($\gamma_1,...,\gamma_{P}, \alpha_1,...,\alpha_{P_f}$).

In our generalization of BHM the spectral degeneracy is calculated
through the steps:

{\em Step 1}: Choice of a reversible protocol of allowed movements in
the space of microstates. Reversible protocol means that each allowed
movement \mbox{${\bf X}_{\rm old} \rightarrow {\bf X}_{\rm new}$}
corresponds to another also allowed movement \mbox{${\bf X}_{\rm new}
\rightarrow {\bf X}_{\rm old}$}. This definition is independent of
the probabilities prescribed to these movements by the particular
dynamics to be used in a practical computer implementation.  These
movements are virtual ones, since they are not actually performed.

{\em Step 2}: To compute the number $N({\bf X},{\bf \Delta s})$ of
allowed movements, for a configuration ${\bf X}$, that changes the
current macrostate ${\bf s}$ to another ${\bf s + \Delta s}$.
Then $\langle N({\bf s}, {\bf \Delta s})\rangle$ is the average of
$N({\bf X}, {\bf \Delta s})$, i.e., the mean number of possible
movements from macrostate ${\bf s}$ to macrostate
${\bf s} + {\bf \Delta s}$;

{\em Step 3}: Due to the reversibility condition (step 1), the total
number of possible movements from macrostate ${\bf s} + {\bf \Delta
s}$ to macrostate ${\bf s}$ is equal to the total number of possible
movements from macrostate ${\bf s}$ to macrostate ${\bf s} + {\bf
\Delta s}$. Thus, we can write down the equation %
\begin{equation}
\label{bhrel}
g({\bf s})\langle N({\bf s}, {\bf \Delta s})\rangle = g({\bf s}+{\bf \Delta s})\langle
N({\bf s}+{\bf \Delta s}, -{\bf \Delta s})\rangle.
\end{equation}
This relation has been shown to be exact for any statistical model
and spectrum degeneracy \cite{pmco98}. Our multiparametric
formulation remain valid based on the same arguments. Therefore,
equation (\ref{bhrel}) can be rewritten as %
\begin{equation}
\ln g({\bf s}+{\bf \Delta s}) - \ln g({\bf s}) =  
\ln \frac{\langle N({\bf s}, {\bf \Delta s})\rangle}{\langle
N({\bf s}+{\bf \Delta s}, -{\bf \Delta s})\rangle}.
\end{equation}
This equation can be iteratively solved for all macrostates ${\bf
s}$, provided $\langle N({\bf s}, {\bf \Delta s})\rangle$ is known
(again, this relation is independent of the way $\langle N({\bf s},
{\bf \Delta s})\rangle$ is obtained). This set of equations is
overdetermined.  However, the spectral degeneracy can be obtained
without solving all equations simultaneously, since all values of
the projections of ${\bf \Delta s}$ along a given direction of the
space of macrostates are equivalent.

The Broad Histogram relation (\ref{bhrel}) is independent of the
procedure by which $\langle N({\bf s}, {\bf \Delta s})\rangle$ is
obtained. For the single parameter case, virtually any procedure can
be adopted in this task as, for instance, a non-biased random walk
\cite{pmco96}, a microcanonical simulation \cite{pmco98b} or the
random walk performed by the Entropic Sampling in the space of
microstates \cite{lima99b}. For the multiparametric case, the choice
of the dynamics is crucial to an efficient implementation, because
the larger the number of parameters, the larger and sparser is the
space of macrostates. In this work we choose a dynamics based on the
Entropic Sampling method (ESM). Lima {\em et al.}  \cite{lima99b}
have shown that this dynamics associated to BHM gives better results
than the traditional Multicanonical calculation. The dynamics is
implemented as follows: first, we implement the classical algorithm
of ESM, performing the visitation in the space of macrostates.
Additionally, after a given number of random walk steps (in our case
one Monte Carlo Step (MCS)), for each visited state ${\bf X}$, we
store the values of $N({\bf X}, {\bf \Delta s})$ cumulatively into
${\bf s}$-histograms. This dynamics is particularly interesting due
to its simplicity. The spectral degeneracy is calculated by the BHM
relation (\ref{bhrel}), instead, avoiding the shortcomings of the
traditional determination of the spectral degeneracy by ESM.

The procedure described above is quite general and can be applied to
any statistical model. As examples of applications of the method we
choose two traditional multiparametric problems: i) The Ising model
with nearest and next-nearest neighbor interactions and ii) the
antiferromagnetic Ising model in an external field. Within the
canonical ensemble, two parameters come from the Hamiltonian and one
(the temperature $T$) from the ensemble definition.

We start describing our results for the Ising model with nearest and
next-nearest neighbor interactions in a $L \times L$ square lattice.
The Hamiltonian of this system is given by
\begin{eqnarray}
  {\cal H} &=& -J_1 \sum_{\rm{nn}}{\sigma_{i}\sigma_{j}} - J_2
  \sum_{\rm{nnn}}{\sigma_{i}\sigma_{j}} \nonumber\\
  &=& -J_1 E_1 - J_2 E_2
\end{eqnarray}
where $\sigma_i=\pm 1$. The summations are carried over the nearest
and next-nearest neighbors, respectively. Each macrostate is defined
by the vector ${\bf s} = (E_1, E_2)$, which is independent of the set
of parameters $J_1$, $J_2$ and $T$. The adopted protocol of movements
is the single spin-flip. In this case, the allowed ${\bf \Delta s}$
are: $(-8,-8)$, $(-8,-4)$, $(-8,0)$, $(-8,4)$, $(-8,8)$, $...$, $(8,
-8)$, $(8,-4)$, $(8,0)$, $(8,4)$, $(8,8)$.  For the implementation of
the random walk in the space of macrostates within ESM, we need to
define the number of entropy updates and the number of Monte Carlo
steps (MCS) between each update. We choose the number of MCS between
successive updates as increasing linearly with the number of updates
already performed, such that when the process is finished, each
macrostate is visited on average $\bar{v}$ times. The number of
updates was chosen as $100$ and $\bar{v}$ as $4000$. Fig. (1)
presents the entropy $S(E_1,E_2)=\ln g(E_1,E_2)$ for a $10 \times 10$
system, obtained by solving (4) from the measured averages $\langle
N({\bf s}, {\bf \Delta s})\rangle$. We have not used any additional
procedure to optimize our simulations, as for example, to use the
entropy obtained from another simulation as initial input for the
Entropic Sampling-based random walk. We also choose to store
informations for all possible values of $E_1$ and $E_2$. Once
$g(E_1,E_2)$ is obtained, we can calculate any thermodynamical
quantity, without resorting again to computer simulations. Of
particular interest is the line defined by the critical temperature
$T_c$ as a function of the ratio
$K=J_2/J_1$. We have calculated the internal energy, magnetization,
specific heat and magnetic susceptibility as {\bf continuous}
functions of both $K$ and $T$. We have used the temperature where the
specific heat presents a maximum as an estimative to the critical
point for each $K$. In figure (2), we present our results for
different lattice sizes. The lines are normalized such that at $K=0$,
$T_c=1$. In the same figure, we present the infinite lattice results
obtained by Fan \& Wu using low-temperature series expansions
\cite{fan69}. Finite size effects are noticed only near to $K=-0.5$
(insets). An important advantage of our method is that it is possible
to blow-up a given region in the space of parameters without need of
additional runs. All points in the critical lines were calculated
from the same $g(E_1, E_2)$ obtained from {\bf only one simulation}
for each size. Traditional approaches \cite{landau80} need a new
simulational run for each pair $(T, K)$. To our knowledge, the
results presented here are the most detailed available. Shteto {\em
et al.} \cite{shteto97} suggest $32 \times 32$ as the limit for their
multiparametric simulations of the Ising model in an external
magnetic field with Entropic Sampling. We have reached this limit
with a much more elaborated problem.

As a second example, we choose to study the Ising model in an
external magnetic field.  The Hamiltonian of this system is given by:
\begin{eqnarray}
  {\cal H} &=& -J\sum_{\rm{nn}}{\sigma_{i}\sigma_{j}} + 
       h \sum_{\rm{i}}{\sigma_{i}} \nonumber \\
  &=& -JE + hm
\end{eqnarray}                                %
where $\sigma_{i}=\pm 1$. The macrostates are defined by ${\bf s}=(E,
m)$.  Fig. (3) presents the entropy $S(E, m)$ for a $10 \times 10$
system. For this model, it is interesting to obtain the critical
temperature as a function of the applied field, for the
antiferromagnetic case ($J<0$). In fig. (4) we present our results.
The theoretical line obtained by Wang and Kim
is also shown \cite{wang97}. The agreement is remarkable for all
values of the field.  Again the pseudo-critical temperature was
determined by the peak position in the specific heat. Due to the
excellent agreement reached already with the sizes we have studied,
we did not simulate larger lattices.

Concerning the CPU time necessary to perform such simulations, we
present in table I the mean visitation on each macrostate, and the
total CPU time. From these results the CPU time $\tau$ to obtain, in
average, $\bar{v}$ visitations on each macrostate scales like $L^a$
where $a \approx 6$ as expected, since a factor $L^{d \times P}$
comes from the volume dependence of the space of macrostates and
$L^d$ from the increase in the time to perform one Monte Carlo step.
Spectral degeneracy calculation methods (as the BHM and
Multicanonical) are not based on thermodynamical concepts and hence
not subjected to critical slowing down, large energy barriers and
other problems. Therefore, it is misleading to compare the CPU time
of our method with other Monte Carlo canonical methods. Additionally,
for traditional methods the larger the system the larger is the number
of simulations at different temperatures one needs in order to
determine the peak position for the specific heat. On the other hand,
within the BHM once one already knows $g(E_1, E_2)$($g(E, m)$), it is
possible to obtain results for any values of $J_1$, $J_2$, $T$ ($J$,
$m$, $T$) {\bf continuously}. Other quantities, as the free energy,
which are hard to obtain by traditional methods, can also be obtained
from only one simulation. In this sense, our method introduces a huge
computational speedup.

In conclusion, we have shown that the Broad Histogram Method can be
extended to obtain spectral degeneracies for multiparametric
Hamiltonians.  This generalization allows us to calculate the
thermodynamical averages for any set of the parameters which define
the model {\bf continuously}, from only one computer run.
Conversely, traditional Monte Carlo simulations need to perform a new
computer run each time any parameter of the system is changed.
Compared with traditional Monte Carlo simulations, this feature of
BHM represents an enormous computational gain. Moreover, the
generalization is such that the method could be applied to any
statistical model.

This work was partially supported by CNPq, CAPES and FAPERJ
(Brazilian Agencies). ARL acknowledge S. L. A. de Queiroz for very
useful discussions.

\newpage
\baselineskip=24pt
\centerline{FIGURE CAPTION}
\bigskip

{\bf Fig. 1 - }{Entropy for a $10 \times 10$ Ising model with nearest and
  next-nearest neighbor interactions. The axes represent the nearest ($E_1$)
  and next-nearest neighbor ($E_2$) terms of the corresponding Hamiltonian.
  The entropy does not cover the whole space because some pairs $(E_1, E_2)$ do not  correspond to accessible states.}

{\bf Fig. 2 - }{Comparison between simulations of the Ising model with nearest
  and next-nearest neighbor interactions (for $L \times L$ lattices) and
  low-temperature expansion \cite{fan69} (solid line). The symbols
  correspond to $L$ equal to: 10 (empty circles), 16 (pluses), 20 (crosses), 32
  (diamonds).  We also show some Metropolis results for L=32 (filled circles).
  Finite size effects are more remarkable near to $K = -0.5$ (upper-left
  inset). In the other inset (down-right) we show the results obtained using
  the ferromagnetic susceptibility to determine the critical point, instead of
  the specific heat.

{\bf Fig. 3 - }{Entropy for a $10 \times 10$ Ising Model in a external
  magnetic field. The axes represent the spin coupling term ($E$) and the
  external field term ($m$).}

{\bf Fig. 4 - }{Comparison between simulation and theoretical results for the
  antiferromagnetic Ising Model in a external magnetic field. The agreement
  between simulation and theoretical results is excellent.}

\newpage
\centerline{TABLE}
\bigskip
\begin{table}[!h]
\nonumber
\begin{center}
\begin{tabular}{|c|c|c|c|}
\hline
Model &  L & Mean Visitation & CPU Time (s) \\ 
\hline
$J_1,J_2$ & 10 & 4945.2 &    1191.6 \\ 
$J_1,J_2$ & 16 & 4537.2 &   22118.2 \\ 
$J_1,J_2$ & 20 & 4425.0 &   86137.8 \\
$J_1,J_2$ & 32 & 4277.4 & 1848197.1 \\
$J, h$    & 10 & 4761.4 &    1082.4 \\ 
$J, h$    & 20 & 4356.8 &   71002.0 \\
\hline
\end{tabular}
\end{center}
\caption{CPU time and mean visitation for each lattice size.
The simulations were carried out on a Dec Alpha 500.}
\end{table}

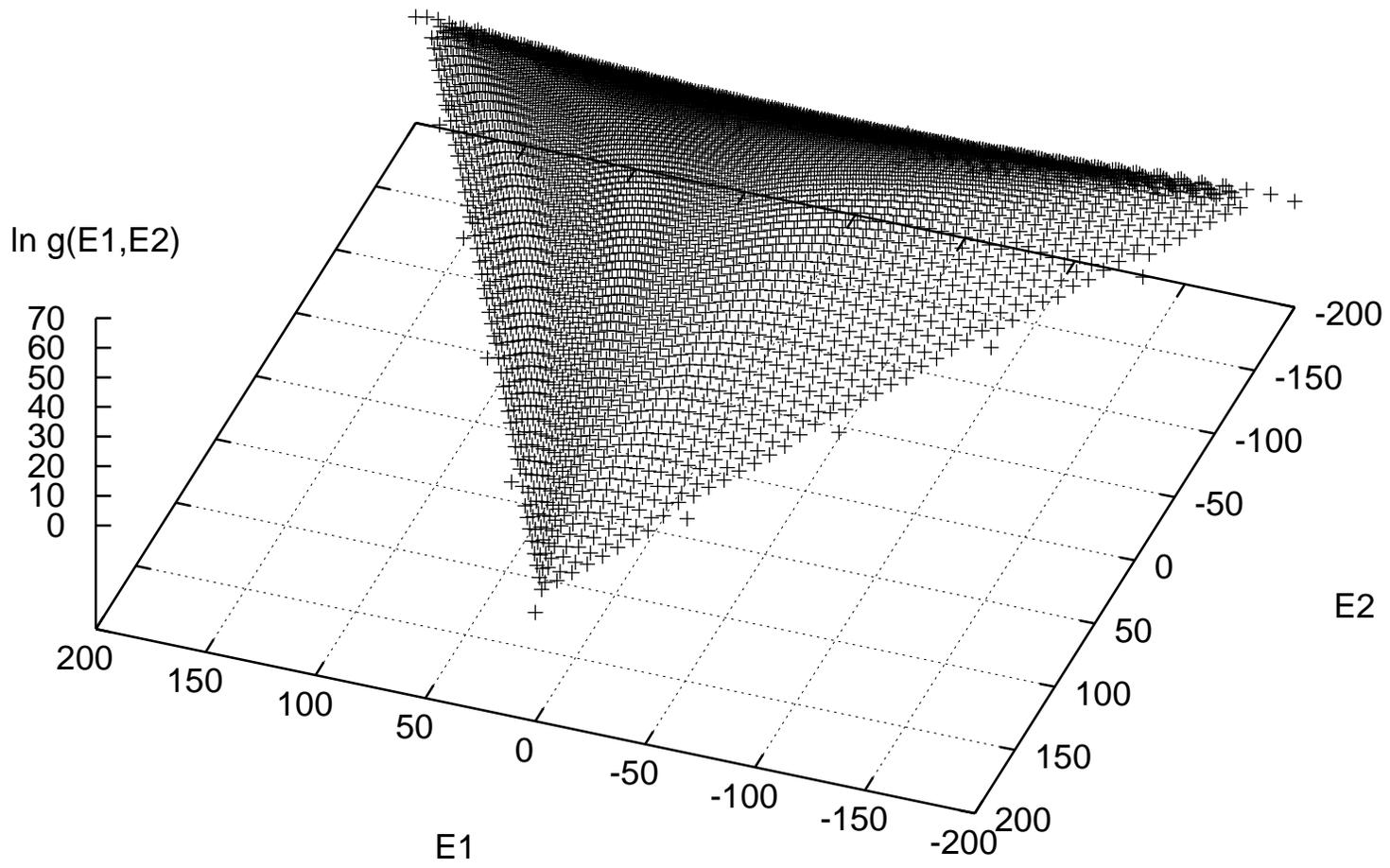
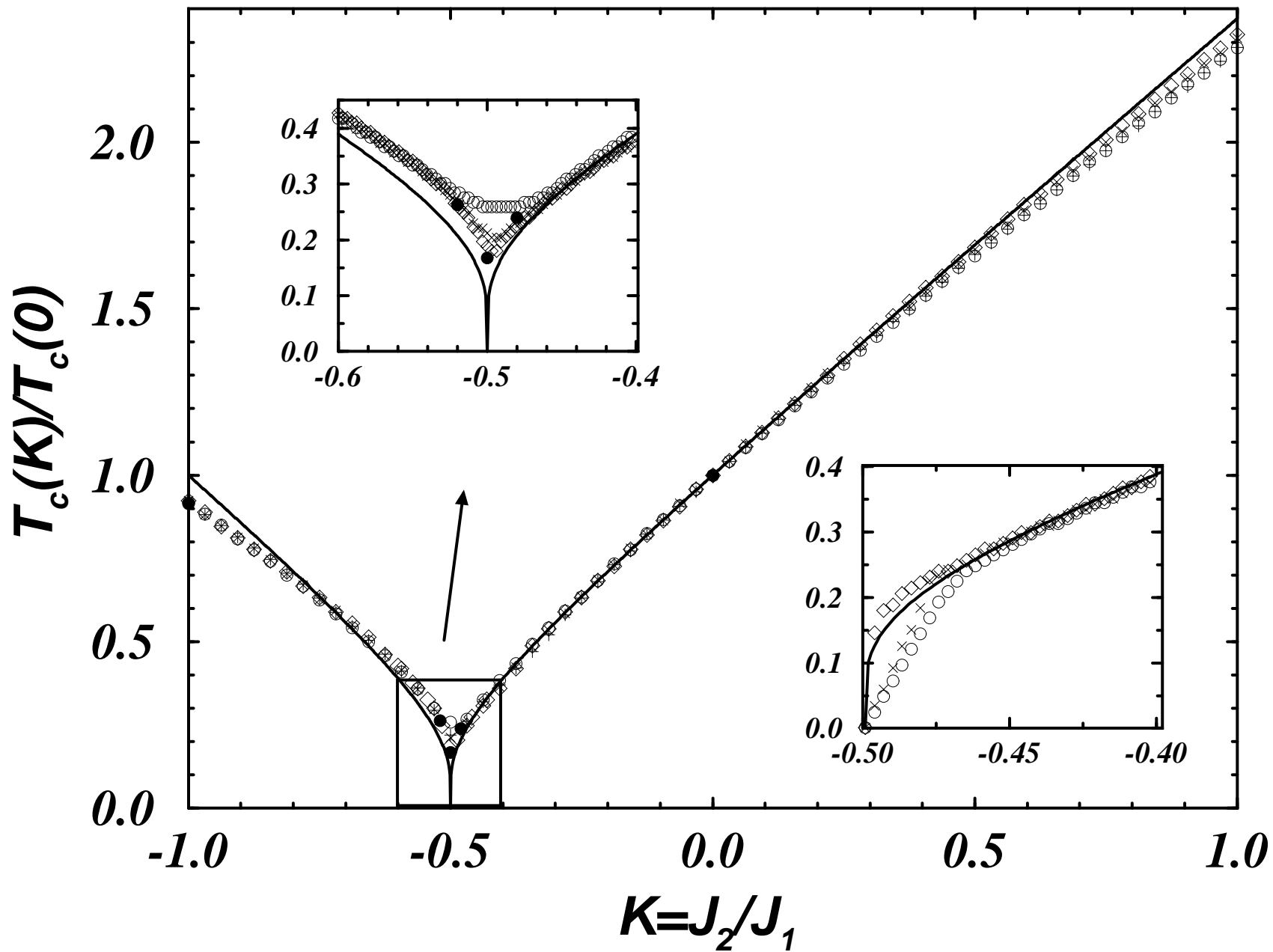
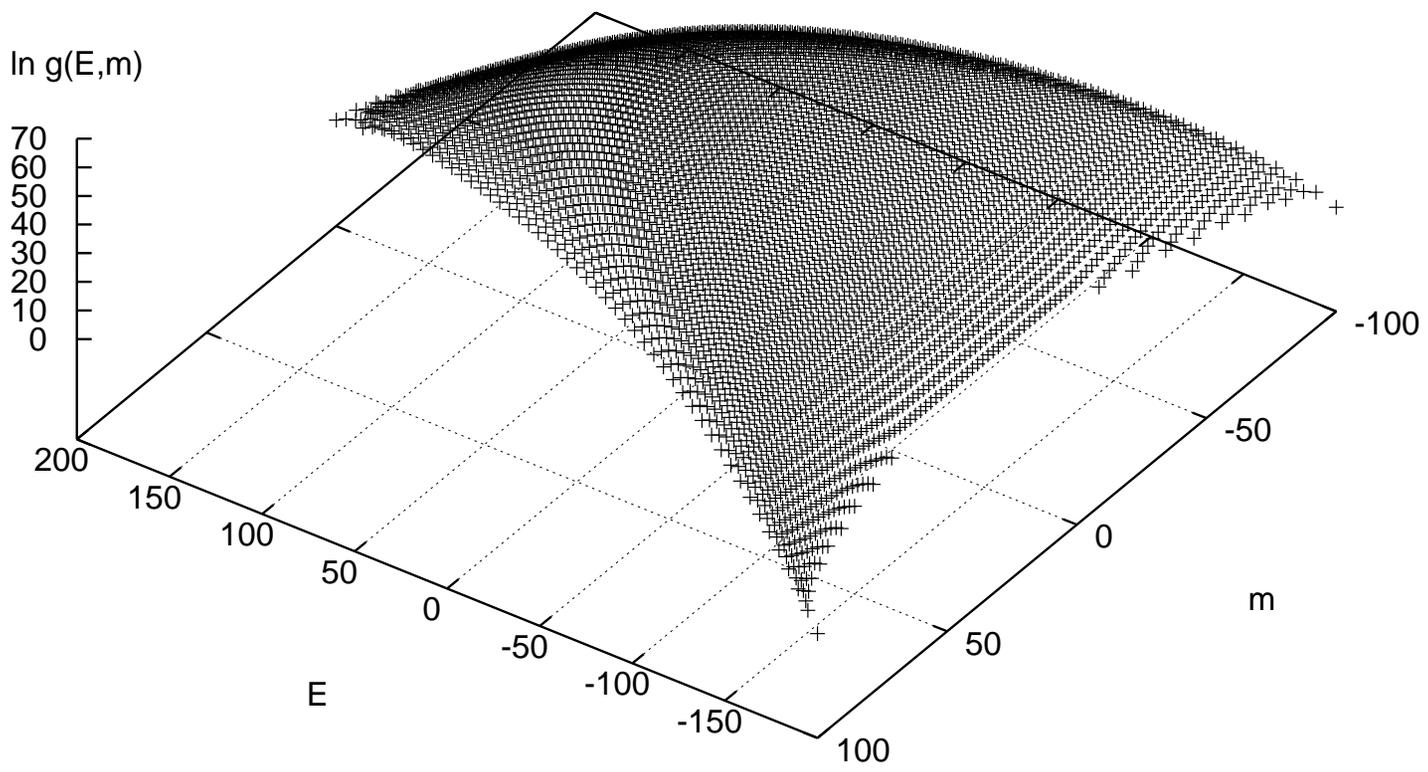
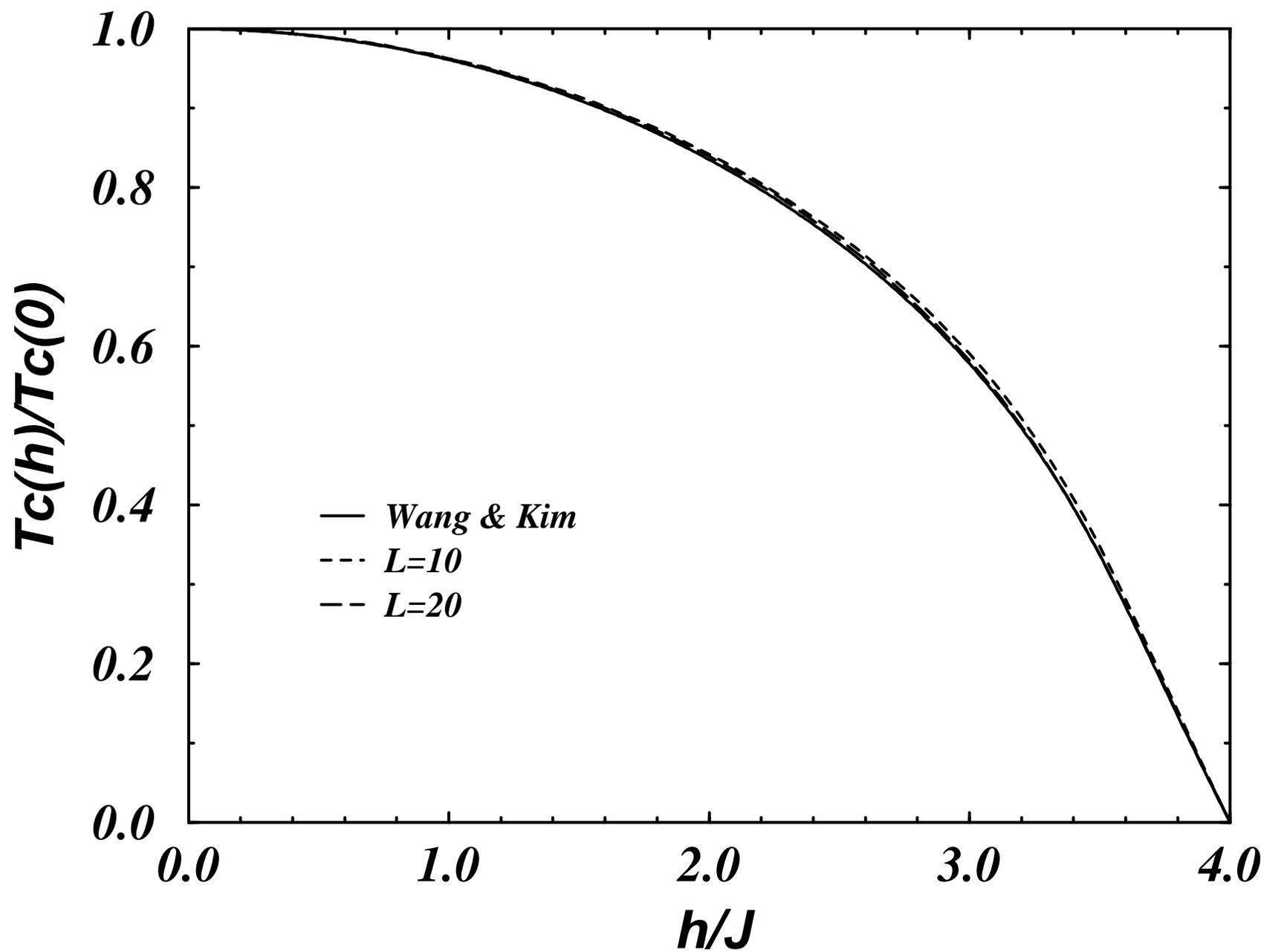
\end{document}